\documentclass[a4paper,11pt]{article}
\pdfoutput=1
\usepackage{pos}
\usepackage{amsmath}
\usepackage{breqn}
\usepackage{braket}
\usepackage{xcolor}
\usepackage{graphicx}
\usepackage{slashed}

\allowdisplaybreaks

\title{Leading Twist-Two Gauge-Variant Counterterms}
\author[1]{Thomas Gehrmann}
\author[2]{Andreas von Manteuffel  }
\author*[1]{Tong-Zhi Yang}

\affiliation[1]{Physik-Institut, Universität Zürich, Winterthurerstrasse 190, 8057 Zürich, Switzerland}
\affiliation[2]{Institut für Theoretische Physik, Universität Regensburg, 93040 Regensburg, Germany}

\emailAdd{thomas.gehrmann@uzh.ch}
\emailAdd{manteuffel@ur.de}
\emailAdd{toyang@physik.uzh.ch}

\abstract{
Anomalous dimensions of twist-two operators govern the scale evolution of parton distribution functions. For off-shell external states, the physical twist-two operators mix with unknown gauge-variant operators under renormalization. In this talk, we apply the method proposed by us in~\cite{Gehrmann:2023ksf} to compute all gauge-variant one-loop counterterm Feynman rules with five legs, which
enter the determination of 
the four-loop splitting functions in QCD.
}
\FullConference{%
  Loops and Legs in Quantum Field Theory - LL2024,\\
  14-19 April, 2024\\
  Wittenberg, Germany
}

\begin{document}
\renewcommand{\hookAfterAbstract}{%
\par\bigskip
\textsc{Report number}:
\href{}{ZU-TH 44/24}
}
\maketitle

\section{Introduction}
QCD predictions for high-energy hadron-collider observables rely on the factorization theorem, which encodes 
the hadron structure in terms of 
universal parton distribution functions (PDFs). The scale evolution of PDFs, described in the DGLAP evolution equations~\cite{Altarelli:1977zs,Gribov:1972ri,Dokshitzer:1977sg}, is governed by splitting functions. Achieving high precision in the determination of PDFs and splitting functions is very desirable for phenomenological applications. The current state-of-the-art calculation for splitting functions extends to the four-loop order, with numerous partial results becoming available recently~\cite{Gracey:1994nn,Gracey:1996ad,Davies:2016jie,Moch:2017uml,Gehrmann:2023cqm,Moch:2021qrk,Falcioni:2023luc,Falcioni:2023vqq,Falcioni:2023tzp,Moch:2023tdj,Gehrmann:2023iah,Basdew-Sharma:2022vya,Falcioni:2024xyt}. These results have already been employed to derive approximate N$^3$LO PDFs~\cite{McGowan:2022nag,NNPDF:2024nan}.

The majority of partial results for 
 four-loop splitting functions (for fixed Mellin moments or specific colour structures)
 are extracted from the computation of 
 off-shell operator matrix elements (OMEs) to four-loop order.  These OMEs are defined as the off-shell matrix elements with a single twist-two quark or gluon 
 operator insertion, and this approach has proven to be computationally efficient in deriving splitting functions. However, under renormalization, the physical twist-two operators mix with unknown gauge-variant (GV) operators. Besides performing the reductions of four-loop integrals for off-shell OMEs with physical twist-two operator insertions, it is the identification of these unknown GV operators or alternatively the determination of the corresponding counterterm Feynman rules that still represents a major challenge in the calculation of the four-loop splitting function. In the past, there have been substantial efforts in this area~\cite{Gross:1974cs,Kluberg-Stern:1974nmx,Kluberg-Stern:1975ebk,Joglekar:1975nu,Collins:1994ee,Hamberg:1991qt}, but a complete solution remained elusive. 
 %Recently, a method was proposed in~\cite{Falcioni:2022fdm} for constructing GV operators with a fixed moment $n$, and this approach has subsequently been extended to determine the leading GV operators and their corresponding Feynman rules for up to five legs with all-$n$ dependence~\cite{Falcioni:2024xav}. 
Generalized BRST and anti-BRST constraints on the form of GV operators
at arbitrary $n$ were recently derived in~\cite{Falcioni:2022fdm}. A method to solve these constraints to compute the corresponding Feynman rules (at leading
order) for up to five legs with all-$n$ dependence was subsequently
presented in~\cite{Falcioni:2024xav}. In~\cite{Gehrmann:2023ksf}, we introduced an alternative method that enables the determination of GV counterterm Feynman rules with all-$n$ dependence. This method has been successfully applied to renormalize physical twist-two operators up to three-loop order~\cite{Gehrmann:2023ksf} in a covariant gauge and to extract the $N_f^2$ contributions~\cite{Gehrmann:2023cqm} to the 
 four-loop pure-singlet splitting functions. In this talk, we further apply this method to derive all leading GV counterterm Feynman rules for five external legs, which constitute a key ingredient to the derivation of 
 the full four-loop splitting functions.

 In Section 2, we provide a brief review of the method described in~\cite{Gehrmann:2023ksf}. Section 3 details the derivations of the leading GV counterterm Feynman rules with five legs. Our results are presented in Section 4. Finally, we conclude in Section 5.

\section{Review of the method for deriving the GV counterterm Feynman rules}
We focus on the renormalization of the  physical twist-two operators in the flavor-singlet sector, specifically
\begin{align}
\label{eq:singletOP}
O^{}_q(n) &= \frac{i^{n-1}}{2}  \bigg[  \bar{\psi}_{i_1} \Delta \cdot \gamma^{} (\Delta \cdot D^{})_{i_1 i_2} ( \Delta \cdot D)^{}_{i_2 i_3}\cdots (\Delta \cdot D)^{}_{i_{n-1} i_n} \psi_{i_n}  \bigg] \,, \nonumber  \\
O^{}_g(n) &=-\frac{i^{n-2}}{2}  \bigg[ \Delta_{\mu_1}  G^{\mu_1}_{a_1 \mu} ( \Delta \cdot D)^{}_{a_1 a_2} \cdots (\Delta \cdot D)^{}_{a_{n-2} a_{n-1}} \Delta_{\mu_n} G^{\mu_n \mu}_{a_{n-1} a_n}  \bigg] \,. 
\end{align}
where $n$ denotes the Mellin moment (spin) of the operators and $\Delta$ is a light-like reference vector with $\Delta^2 =0$. As usual, the symbol $\psi$ represents the quark field and $G$ denotes the gluon field strength tensor. The covariant derivative is given by $D^\mu_{} = \partial_\mu \, \delta_{} -i g_s  \boldsymbol{T}{}^a  A^a_\mu $, where $\boldsymbol{T}{}^a $ are the generators of the gauge group and $A^a_\mu$ is the gauge field.
As explained in~\cite{Gehrmann:2023ksf,Gehrmann:2022euk}, a naive renormalization 
\begin{align}
\label{eq:mixingOqOg}
\left( \begin{array}{c} 
O_q \\
O_g
\end{array} \right)^{\text{R,naive}} =  \left( \begin{array}{cc} 
Z_{qq} & Z_{qg} \\
Z_{gq} & Z_{gg} 
\end{array} \right)   \left( \begin{array}{c} 
O_q \\
O_g 
\end{array} \right)^{\text{B}}\,
\end{align}
is not enough to renormalize the physical twist-two operators beyond one-loop order. Instead, the renormalization should be extended to the following form:
\begin{align}
\label{eq:mixingOqOgOA}
\left( \begin{array}{c} 
O_q \\
O_g \\
O_{ABC}\\
\end{array} \right)^{\text{R}} =  \left( \begin{array}{ccc} 
Z_{qq} & Z_{qg} & Z_{qA} \\
Z_{gq} & Z_{gg} & Z_{gA}  \\ 
0 & 0 & Z_{AA}  \\
\end{array} \right)   \left( \begin{array}{c} 
O_q \\
O_g \\
O_{ABC}\\
\end{array} \right)^{\text{B}} + 
\left( 
\begin{array}{c}
\left[Z O\right]_{q}^{\textrm{GV}}\\
\left[Z O\right]_{g}^{\textrm{GV}}\\
\left[Z O\right]_{A}^{\textrm{GV}}
\end{array}
\right)^{\text{B}} ,
\end{align}
where we use the shorthand notation $O_{ABC} = O_A+ O_B +O_C$, with $O_A,\, O_B,\,O_C$ denoting the leading GV twist-two operators involving all-gluon, quark-gluon, and ghost-gluon fields, respectively. At higher orders of the strong coupling, three GV counterterms denoted as $\left[Z O\right]_{q}^{\textrm{GV}},
\left[Z O\right]_{g}^{\textrm{GV}},
\left[Z O\right]_{A}^{\textrm{GV}}$ are required in addition to renormalize the physical operators. For the GV counterterms, $Z$ and $O$ are combined as $\left[ZO\right]$ 
since it is in general not possible (and also not required in practice) 
 to separate the renormalization constants $Z$ from their associated operators $O$ when retaining the all-$n$ dependence in our approach.

One can formally expand the GV counterterms in the following form:
\begin{align}
    \label{eq:CODec}
    \left[Z O\right]_{i}^{\textrm{GV}}  = \sum_{l=2}^\infty a_s^{l} \left[Z O\right]_{i}^{\textrm{GV},\,\left(l\right)}, \text{with } i=q,g, A \,,
\end{align} 
with $a_s = \alpha_s/(4 \pi)$. The contributions from GV operators or counterterms start at different orders of $a_s$,
\begin{align}
\label{eq:powerCounting}
    Z_{qA} = \mathcal{O}(a_s^2), \quad Z_{gA} =& \mathcal{O}(a_s), \quad \left[Z O\right]_{q}^{\textrm{GV}}= \mathcal{O}(a_s^3), \quad \left[Z O\right]_{g}^{\textrm{GV}} = \mathcal{O}(a_s^2), \nonumber \\
    & Z_{AA} = \mathcal{O}(a_s^0),\quad \left[ZO\right]_{A}^{\textrm{GV}} = \mathcal{O}(a_s) \,.
\end{align} 
An interesting observation in~\cite{Gehrmann:2023ksf} is that the (counterterm) Feynman rules for the corresponding GV operators can be extracted by inserting eq.~\eqref{eq:mixingOqOgOA} into matrix elements with more than two off-shell external states, even without knowing the operators themselves. Without loss of generality, we will consider the renormalization of the physical gluon operator:
\begin{align}
    O^{\text{R}}_g = Z_{gq} O^{\text{B}}_q + Z_{gg} O^{\text{B}}_g + Z_{gA} O_{ABC}^{\text{B}}+ \left[Z O\right]_{g}^{\textrm{GV}}\,.
\end{align}
Since we are discussing the renormalization of a leading-twist operator, it suffices to consider one-particle-irreducible (1PI) OMEs with all-off-shell external states composed of two particles of type $j$ plus $m$ gluons,
\begin{align}
\label{eq:2jMgluonState} 
&\braket{j|O_g|j+ m\,g}^{\mu_1\cdots\mu_m,\,\text{R}}_{\text{1PI}} =  Z_j (\sqrt{Z_g})^m \bigg[  \braket{j| (Z_{gq} O_q+ Z_{gg} O_g  ) |j+m \,g} ^{\mu_1\cdots\mu_m,\,\text{B}}_{ \text{1PI}} \bigg]  \nonumber  \\
 & \qquad + Z_j (\sqrt{Z_g})^m \bigg[  Z_{gA}  \braket{j|O_{ABC}|j+m\,g} ^{\mu_1\cdots\mu_m,\,\text{B}}_{ \text{1PI}} + \braket{j|\left[Z O\right]_{g}^{\textrm{GV}}|j+ m\,g }^{\mu_1\cdots\mu_m,\,\text{B}}_{ \text{1PI}}  \bigg]\,,
\end{align} 
where $j$ can be a quark($q$), gluon($g$), or ghost($c$) external state, and $\sqrt{Z_j}$ is the corresponding wave function renormalization constant. To continue, we expand the off-shell OMEs according to the number of loops $l$ and legs $m+2$, 
\begin{align}
 \braket{j|O|j+ m\,g}^{\mu_1\cdots\mu_m\text{}} = \sum_{l=0}^{\infty}\left[ \braket{j|O|j+ m \,g}^{\mu_1\cdots\mu_m, \,(l),\,(m)\text{}}_{\text{}} \right] a_s^l  \, g_s^m \,.
\end{align} 
Since the left-hand side of equation~\eqref{eq:2jMgluonState} is ultraviolet-renormalized and infrared finite (with all external states being off-shell), the summation on the right-hand side of the equation must also be free of poles in the dimensional regulator $\epsilon$. This provides a method for deriving the GV (counterterm) Feynman rules by computing the corresponding off-shell OMEs order by order in the strong coupling. As an example, the two-ghost plus $m$-gluon Feynman rules for $O_C$ operator can be written in the following compact form, 
\begin{align}
\label{eq:formulaOC}
  \textcolor{black}{{\braket{c|O_C|c+ m\,g}^{\mu_1\cdots\mu_m, \,\textcolor{black}{(0)},\,(m)}_{\text{}}=\frac{-1}{Z_{gA}^{(1)}} \left[\braket{c|O_g|c+ m\,g}^{\mu_1\cdots\mu_m, \,(1),\,(m),\,\text{B}}_{\text{1PI}} \right]_{\text{div}} }} \,.
\end{align}
Here the subscript 'div' is the pole part in $\epsilon$, and $Z_{gA}^{(1)}$ is the leading GV renormalization constant
\begin{equation}
\label{eq:leadingZgA}
    Z_{gA}^{(1)} = \frac{1}{\epsilon} \frac{C_A}{n(n-1)}\,.
\end{equation}
 Note that the above method is general and can be applied to extract (counterterm) Feynman rules to any number of loops and legs. In the following, we focus on the determination of Feynman rules with five legs for the leading GV operators, i.e.\ operators $O_{A}\,,O_B\,, O_C$.

\section{Computations of five-leg Feynman rules for operators $O_A, O_B, O_C$}
\begin{figure}
\begin{center}
\begin{minipage}{4.6cm}
\includegraphics[scale=0.9]{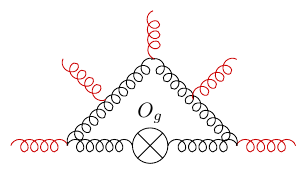} 
\end{minipage}
\begin{minipage}{4.6cm}
\includegraphics[scale=0.9]{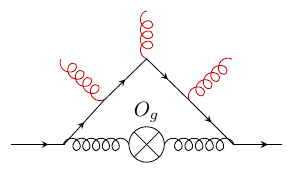} 
\end{minipage}
\begin{minipage}{4.6cm}
\includegraphics[scale=0.9]{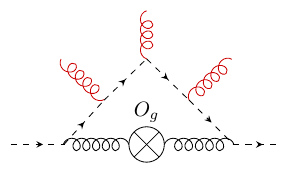} 
\end{minipage}
\caption{Sample 1-loop diagrams to determine the Feynman rules with 5 legs stemming from $O_A, O_B, O_C$ operators respectively. Here, all external states are off-shell, and all diagrams involve the insertion of the physical gluon operator $O_g$.}
\label{fig:OneloopFiveLegs}
\end{center}
\end{figure}

To extract Feynman rules for operators $O_A\,,O_B\,,O_C$, we can simply expand the eq.~\eqref{eq:2jMgluonState} to one-loop order, where the two-loop GV counterterms are absent. In the following, we focus on the case with five legs. Some example Feynman diagrams are shown in Fig.~\ref{fig:OneloopFiveLegs}.

We consider the computations of one-loop all-off-shell five-particle scatterings with an insertion of the operator $O_g$, involving 14 scales
\begin{align}
    p_1^2, p_2^2,  p_3^2, p_4^2, p_1 \cdot p_2, p_1 \cdot p_3, p_1 \cdot p_4, p_2 \cdot p_3,  p_2 \cdot p_4,p_3 \cdot p_4, \Delta \cdot p_1, \Delta \cdot p_2,  \Delta \cdot p_3, \Delta \cdot p_4\,. 
\end{align}
Here, we eliminated $p_5$ by momentum conservation 
\begin{align}
    \sum_{i=1}^5 p_i =0 \,.
\end{align}

To retain the all-$n$ dependence for the off-shell OMEs, we employ the generating function method introduced in~\cite{Ablinger:2012qm,Ablinger:2014nga}. This method sums non-standard terms in the Feynman rules, such as $(\Delta \cdot p)^{n-1}$ into linear propagators that depend on a tracing parameter, $t$. For example, \begin{align} 
\label{eq:tracingT} ( \Delta \cdot p)^{n-1} \to \sum_{n=1}^\infty ( \Delta \cdot p)^{n-1} t^n = \frac{t}{1- t \Delta \cdot p}\,. 
\end{align}
By working in $t$-space, we can perform standard IBP reductions~\cite{Chetyrkin:1981qh,Laporta:2000dsw} and extract the desired $n$-space results in the final step by expanding the parameter $t$ around $t = 0$.

It is highly non-trivial to directly compute the one-loop all-off-shell OMEs with 14 scales. Luckily the computation can be greatly simplified by analyzing the structures of Feynman rules for a twist-two operator. According to the dimensional analysis shown in~\cite{Gehrmann:2023ksf}, the Feynman rules for operators $O_B$ and $O_C$ consist of one Lorentz tensor only
\begin{align}
\label{eq:ansatzForOBOC}
&\left[\braket{c|O_C|c+ m\,g}^{\mu_1\cdots\mu_m, \,(0),\,(m) }_{\text{}} \right]  = c_m \Delta^{\mu_1} \Delta^{\mu_2} \cdots \Delta^{\mu_m} \,, \nonumber \\
& \left[\braket{q|O_B|q+ m\,g}^{\mu_1\cdots\mu_m, \,(0),\,(m)}_{\text{}} \right]  = b_m \Delta^{\mu_1} \Delta^{\mu_2} \cdots \Delta^{\mu_m} \,,
\end{align}
and the coefficients $c_m\,,b_m$ are functions of $\Delta \cdot p_i$ only. The Feynman rule for operator $O_A$ with five legs is composed of 31 tensor structures:
\begin{align}
\label{eq:5gAnsatz}
 &\left[\braket{g|O_A|g g gg}^{\mu_1\mu_2\mu_3\mu_4 \mu_5,\,(0),\,(3)}_{\text{}} \right] = a_1  \Delta^{\mu_1} \Delta^{\mu_2} \Delta^{\mu_3}  \Delta^{\mu_4} \Delta^{\mu_5} \nonumber \\
 & \quad +  \Delta^{\mu_1} \Delta^{\mu_2} \Delta^{\mu_3} \Delta^{\mu_4} ( a_2 p_1^{\mu_5} + a_3 p_2^{\mu_5} + a_4 p_3^{\mu_5} + a_5 p_4^{\mu_5} ) \nonumber \\
 & \quad + \Delta^{\mu_1} \Delta^{\mu_2} \Delta^{\mu_3} \Delta^{\mu_5} ( a_6 p_1^{\mu_4} + a_7 p_2^{\mu_4} + a_8 p_3^{\mu_4} + a_9 p_4^{\mu_4} ) \nonumber \\
 &\quad + \Delta^{\mu_1} \Delta^{\mu_2} \Delta^{\mu_4} \Delta^{\mu_5} ( a_{10} p_1^{\mu_3} + a_{11} p_2^{\mu_3} + a_{12} p_3^{\mu_3} + a_{13} p_4^{\mu_3} ) \nonumber \\
 & \quad  + \Delta^{\mu_1} \Delta^{\mu_3} \Delta^{\mu_4} \Delta^{\mu_5} ( a_{14} p_1^{\mu_2} + a_{15} p_2^{\mu_2} + a_{16} p_3^{\mu_2} + a_{17} p_4^{\mu_2} ) \nonumber \\
 & \quad + \Delta^{\mu_2} \Delta^{\mu_3} \Delta^{\mu_4} \Delta^{\mu_5} ( a_{18} p_1^{\mu_1} + a_{19} p_2^{\mu_1} + a_{20} p_3^{\mu_1} + a_{21} p_4^{\mu_1} ) \nonumber \\
 & \quad +a_{22} \Delta^{\mu_1} \Delta^{\mu_2} \Delta^{\mu_3} g^{\mu_4 \mu_5}+a_{23} \Delta^{\mu_1} \Delta^{\mu_2} \Delta^{\mu_4} g^{\mu_3 \mu_5} +a_{24} \Delta^{\mu_1} \Delta^{\mu_2} \Delta^{\mu_5} g^{\mu_3 \mu_4} +a_{25} \Delta^{\mu_1} \Delta^{\mu_3} \Delta^{\mu_4} g^{\mu_2 \mu_5} \nonumber \\
 &\quad +a_{26} \Delta^{\mu_1} \Delta^{\mu_3} \Delta^{\mu_5} g^{\mu_2 \mu_4} +a_{27} \Delta^{\mu_1} \Delta^{\mu_4} \Delta^{\mu_5} g^{\mu_2 \mu_3} +a_{28} \Delta^{\mu_2} \Delta^{\mu_3} \Delta^{\mu_4} g^{\mu_1 \mu_5} \nonumber \\
 & \quad +a_{29} \Delta^{\mu_2} \Delta^{\mu_3} \Delta^{\mu_5} g^{\mu_1 \mu_4} +a_{30} \Delta^{\mu_2} \Delta^{\mu_4} \Delta^{\mu_5} g^{\mu_1 \mu_3} +a_{31} \Delta^{\mu_3} \Delta^{\mu_4} \Delta^{\mu_5} g^{\mu_1 \mu_2}\,,
\end{align}
where the coefficient $a_1$ is linear in the Mandelstam variables $p_i^2, p_i \cdot p_j$. Similarly to $c_m$ and $d_m$ in eq.~\eqref{eq:ansatzForOBOC}, the coefficients $a_j$ for $j\geq 2$ depend solely on $\Delta \cdot p_i$.

Owing to these properties, the evaluation of the 
five-parton one-loop OME and the corresponding IBP reduction can be performed by setting all Mandelstam variables to non-zero rational numbers, which significantly simplifies the workflow. A single numerical sample is sufficient to determine the Feynman rules involving quarks or ghosts, with an additional sample used for cross-checking. In the case of the five-gluon Feynman rules, one numerical sample is also enough to fix the unknown parameters, thanks to the symmetry constraints imposed by Bose statistics. 

Another simplification arises from the fact that the Feynman rules are contained in the $\epsilon$-divergent part of the OMEs. Consequently, only the following two types of Feynman integrals contribute: \begin{align} I_1 &= (\mu^2)^\epsilon \int \frac{d^d l}{i \pi^{d/2}} \frac{1}{(l-q_1)^2 l^2} \,, \nonumber \\
I_2 &= (\mu^2)^\epsilon \int \frac{d^d l}{i \pi^{d/2}} \frac{1}{(l-q_1)^2 l^2 \big(1 - t \Delta \cdot (l+q_2) \big) } \,, \end{align} 
where $q_1\,,\,q_2$ are linear combinations of the momenta $p_1\,,\,\cdots\,,\, p_{4}$ and $\mu$ is the 't Hooft mass. These two master integrals can be solved to high orders in $\epsilon$ without much effort. Here, we only need their single-pole terms:
\begin{align}
&I_1 = \frac{1}{\epsilon} + \mathcal{O}(\epsilon^0)\,, \nonumber \\
&I_2 = \frac{1}{\epsilon} \left[ \frac{\ln (1-t \Delta \cdot q_1  - t \Delta \cdot q_2) - \ln (1-t \Delta \cdot q_2)  }{- t \Delta \cdot q_1}  \right] + \mathcal{O}(\epsilon^0)\,.
\end{align} 

With these simplifications, the computational steps follow a standard chain, where several packages are involved, including
\texttt{QGRAF}~\cite{Nogueira:1991ex}, \texttt{FORM}~\cite{Vermaseren:2000nd},
 \texttt{Apart}~\cite{Feng:2012iq},
\texttt{Kira}~\cite{Klappert:2020nbg},
\texttt{FiniteFlow}~\cite{Peraro:2019svx}, \texttt{MultivariateApart}~\cite{Heller:2021qkz}. We 
obtain the $t$-space results for the single-pole contributions in $\epsilon$ of one-loop off-shell OMEs. These results are then transformed back into $n$-space. Next, we perform renormalization to one-loop order according to eq.~\eqref{eq:2jMgluonState}, which provides us with the Feynman rules with five legs for the operators $O_{A},\,O_B,\, O_C$.

\section{Results}
We are now ready to present the all-$n$ Feynman rules for the $O_B,\, O_C,\, O_A$ operators with 5 legs. The Feynman rules up to four legs within the same framework can be found in ~\cite{Gehrmann:2023ksf}. Using the convention of all momenta flowing into the vertices, the results read
\begin{align}
\label{eq:OBFeynmanrules} \nonumber \\[.5ex]
& \includegraphics[scale=1.0]{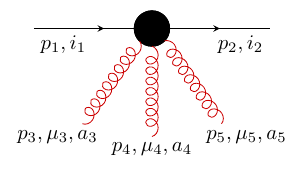}\nonumber\\
&\quad \to   \frac{-g_s^3}{4} \frac{1+(-1)^n}{2 } \slashed{\Delta}_{}  \Delta ^{\mu _3} \Delta ^{\mu _4} \Delta^{\mu_5}
   \left( T^{a } \right)_{i_2 i_1} 
 \sum_{j_1=0}^{n-4} \sum_{j_2=0}^{j_1} \bigg(
     \left(\Delta \cdot p_4\right){}^{j_2} \left(\Delta \cdot
   \left(p_4+p_5\right)\right){}^{-j_1+n-4}  \nonumber\\ 
     &\qquad 
     \times \Big[  
     \left(5 \text{Tr}(T_A^a T_A^{a_3} T_A^{a_5} T_A^{a_4}) -4 \text{Tr}(T_A^a T_A^{a_3} T_A^{a_4} T_A^{a_5}) \right) \left(-\Delta \cdot
   \left(p_1+p_2\right)\right){}^{j_1-j_2}\nonumber \\
   & \qquad -\left( \text{Tr}(T_A^a T_A^{a_3} T_A^{a_4} T_A^{a_5}) -2 \text{Tr}(T_A^a T_A^{a_3} T_A^{a_5} T_A^{a_4}) \right) \left(-\Delta \cdot p_3\right){}^{j_1-j_2} 
     \Big] \bigg)   \nonumber \\
     & \qquad + \text{permutations of three gluons}\,, 
\end{align}
\begin{align}
\label{eq:OCFeynmanrules} \nonumber \\[.5ex]
& \includegraphics[scale=1.0]{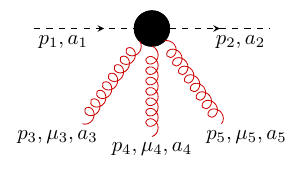}\nonumber\\
&\quad \to \Bigg[ \frac{i g_s^3}{192} \frac{1+(-1)^n}{2 } \Delta^{\mu_3} \Delta^{\mu_4} \Delta^{\mu_5}  \bigg\{\frac{3}{2} \left(f_{11}-2 f_{12}\right) \sum_{j_1=0}^{n-3} \left(\left(\Delta \cdot p_4\right){}^{j_1} \left(-\Delta \cdot
   p_3\right){}^{-j_1+n-3}\right) \nonumber \\
   & \qquad +12 f_8 \sum_{j_1=0}^{n-3} \left(\left(\Delta \cdot p_4\right){}^{j_1} \left(\Delta \cdot
   \left(-p_1-p_3\right)\right){}^{-j_1+n-3}\right) \nonumber \\
   & \qquad +4 \left(6 f_2+13 f_8+f_9\right) \sum_{j_1=0}^{n-3} \sum_{j_2=0}^{j_1} \left(\left(\Delta \cdot
   \left(p_1+p_3\right)\right){}^{j_1-j_2} \left(\Delta \cdot \left(-p_2-p_5\right)\right){}^{j_2} \left(-\Delta \cdot
   p_2\right){}^{-j_1+n-3}\right) \nonumber \\
   & \qquad  +4 \left(6 f_2+4 f_8+f_9\right) \sum_{j_1=0}^{n-3} \sum_{j_2=0}^{j_1} \left(\left(\Delta \cdot p_4\right){}^{j_1-j_2}
   \left(\Delta \cdot \left(-p_2-p_5\right)\right){}^{j_2} \left(-\Delta \cdot p_2\right){}^{-j_1+n-3}\right)  \nonumber \\
   & \qquad -3 \left(6
   f_2+6 f_3+6 f_4+6 f_5+6 f_6+f_8-f_9-f_{10}\right) \nonumber \\
   & \qquad \times  \sum_{j_1=0}^{n-3} \sum_{j_2=0}^{j_1} 
 \left(\left(-\Delta \cdot p_3\right){}^{j_1-j_2}
   \left(\Delta \cdot \left(-p_3-p_5\right)\right){}^{j_2} \left(\Delta \cdot
   \left(p_1+p_2\right)\right){}^{-j_1+n-3}\right) \nonumber \\
   & \qquad +\left(-6 f_1-14 f_7+13 f_{11}\right) \Delta \cdot (p_1-p_2) \nonumber \\
   & \qquad \times  \sum_{j_1=0}^{n-4} \sum_{j_2=0}^{j_1} \left(\left(-\Delta \cdot p_3\right){}^{j_1-j_2} \left(-\Delta \cdot
   \left(p_3+p_5\right)\right){}^{j_2} \left(\Delta \cdot \left(p_1+p_2\right)\right){}^{-j_1+n-4}\right)\nonumber \\
   & \qquad -4 \left(6
   f_2+f_8+4 f_9\right)  \sum_{j_1=0}^{n-3} \sum_{j_2=0}^{j_1} \left(\left(-\Delta \cdot p_4\right){}^{j_2} \left(\Delta \cdot
   \left(-p_4-p_5\right)\right){}^{j_1-j_2} \left(\Delta \cdot p_2\right){}^{-j_1+n-3}\right)\nonumber \\ 
   & \qquad +\left(-6 f_1+f_7+4
   f_{11}\right) \Delta \cdot (p_1-p_2)   \nonumber \\
   & \qquad \times \sum_{j_1=0}^{n-4} \sum_{j_2=0}^{j_1} \left(\left(-\Delta \cdot
   \left(p_3+p_4\right)\right){}^{j_1-j_2} \left(\Delta \cdot p_5\right){}^{j_2} \left(-\Delta \cdot
   p_3\right){}^{-j_1+n-4}\right) \nonumber \\
   & \qquad +18 f_8 \sum_{j_1=0}^{n-3} \left(\left(\Delta \cdot \left(p_2+p_5\right)\right){}^{j_1}
   \left(\Delta \cdot \left(-p_1-p_3\right)\right){}^{-j_1+n-3}\right) \nonumber \\
   & \qquad -3 e_1 \sum_{j_1=0}^{n-3} \sum_{j_2=0}^{j_1} \sum_{j_3=0}^{j_2} \left(\left(\Delta \cdot p_2\right){}^{j_1-j_2} \left(-\Delta \cdot
   \left(p_1+p_3\right)\right){}^{j_2-j_3} \left(\Delta \cdot \left(p_2+p_5\right)\right){}^{j_3} \left(-\Delta \cdot
   p_1\right){}^{-j_1+n-3}\right) \nonumber \\
   & \qquad -9 \left(f_7-2 f_9\right) \sum_{j_1=0}^{n-3}\left(\left(\Delta \cdot p_5\right){}^{j_1}
   \left(\Delta \cdot \left(p_4+p_5\right)\right){}^{-j_1+n-3}\right)\bigg\} + (p_1\leftrightarrow p_2,\,a_1\leftrightarrow a_2) \Bigg] \nonumber \\
   & \qquad + \text{permutations of three gluons}\,,  
\end{align}
\begin{align}
\label{eq:OAFeynmanrules} \nonumber \\[.5ex]
& \includegraphics[scale=1.0]{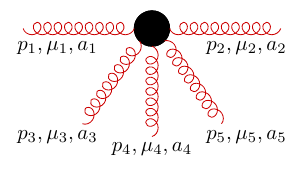}\nonumber\\
&\quad \to i g_s^3 \frac{1+(-1)^n}{2}\bigg\{ -\frac{1}{16} \left(2 f_8-f_9\right)  \Delta^{\mu _1} \Delta^{\mu _2} \Delta^{\mu _3} g^{\mu _4 \mu _5}
   \bigg[\sum_{j_1=0}^{n-3} \left(\left(-\Delta \cdot p_3\right){}^{j_1} \left(\Delta \cdot p_1\right){}^{-j_1+n-3}\right)\nonumber \\
   & \qquad +6
   \sum_{j_1=0}^{n-3}  \left(\left(\Delta \cdot \left(p_1+p_3\right)\right){}^{j_1} \left(\Delta \cdot
   p_1\right){}^{-j_1+n-3}\right)\bigg] \nonumber \\
   &\qquad+\frac{1}{48}  \Delta^{\mu _2} \Delta^{\mu _4} \Delta^{\mu _5}
   \left(2 g^{\mu _1 \mu _3} \Delta \cdot p_1+\Delta^{\mu _3} p_1^{\mu _1}-4 \Delta^{\mu _1} p_1^{\mu _3}\right) \nonumber \\
   & \qquad \times
   \bigg[\left(6 f_2+13 f_8+f_9\right) \sum_{j_1=0}^{n-4} \sum_{j_2=0}^{j_1} \left(\left(\Delta \cdot \left(p_1+p_3\right)\right){}^{j_1-j_2}
   \left(\Delta \cdot \left(-p_2-p_5\right)\right){}^{j_2} \left(-\Delta \cdot p_2\right){}^{-j_1+n-4}\right) \nonumber \\
   & \qquad +\left(6
   f_2+4 f_8+f_9\right) \sum_{j_1=0}^{n-4} \sum_{j_2=0}^{j_1} \left(\left(\Delta \cdot p_4\right){}^{j_1-j_2} \left(\Delta \cdot
   \left(-p_2-p_5\right)\right){}^{j_2} \left(-\Delta \cdot p_2\right){}^{-j_1+n-4}\right)\bigg] \nonumber \\
   &\qquad -\frac{1}{64} e_1 \Delta^{\mu _3} \Delta^{\mu _4} \Delta^{\mu _5} \left(\Delta^{\mu _2} p_1^{\mu _1}-\Delta^{\mu _1}
   p_2^{\mu _2}\right) \nonumber \\
   & \qquad \times  \sum_{j_1=0}^{n-4} \sum_{j_2=0}^{j_1} \sum_{j_3=0}^{j_2} \left(\left(\Delta \cdot p_2\right){}^{j_1-j_2} \left(\Delta \cdot
   \left(-p_1-p_3\right)\right){}^{j_2-j_3} \left(\Delta \cdot \left(p_2+p_5\right)\right){}^{j_3} \left(-\Delta \cdot
   p_1\right){}^{-j_1+n-4}\right)\nonumber \\
   & \qquad + \Delta^{\mu _1} \Delta^{\mu _2} \Delta^{\mu _3} \Delta^{\mu _4} \Delta^{\mu _5} \bigg[\left(\frac{1}{192} p_3\cdot p_3 (-3 e_1- e_2 )+\frac{1}{384} p_4\cdot p_4 (e_1+e_3)-\frac{1}{64} e_1
   p_1\cdot p_1\right) \nonumber \\
   & \qquad \times \sum_{j_1=0}^{n-5} \sum_{j_2=0}^{j_1} \sum_{j_3=0}^{j_2}  \left(\left(\Delta \cdot p_2\right){}^{j_1-j_2} \left(\Delta \cdot
   \left(-p_1-p_3\right)\right){}^{j_2-j_3} \left(\Delta \cdot \left(p_2+p_5\right)\right){}^{j_3} \left(-\Delta \cdot
   p_1\right){}^{-j_1+n-5}\right) \nonumber \\
   &\qquad +  \left( p_1\leftrightarrow p_2, \, p_3\leftrightarrow p_5,\,  a_1\leftrightarrow a_2, \, a_3\leftrightarrow a_5 \right) \bigg]  \bigg\}  + \text{permutations of five gluons} \,,  
\end{align}
where we have defined the following linearly independent color structures for operators $O_A$ and $O_C$:
\begin{align}
   &f_1 = d_{4 f}^{a_3a_4a_5a_1a_2},\, f_2 =d_{4 f}^{a_2a_4a_5a_1a_3},\,f_3 =d_{4
   f}^{a_2a_3a_4a_1a_5},\, \nonumber \\
   & f_4= d_{4 f}^{a_1a_4a_5a_2a_3},\,f_5 = d_{4
   f}^{a_1a_3a_4a_2a_5},\,f_6 =d_{4
   f}^{a_1a_2a_4a_3a_5},\, \nonumber \\
   &f_7= f^{a a_4a_5}\
   f^{aba_3}\ f^{ba_1a_2},\,f_8 = f^{aa_2a_5}\
   f^{aba_4}\ f^{ba_1a_3},\,f_9 = f^{aa_4a_5}\
   f^{aba_2}\ f^{ba_1a_3},\,\nonumber \\
   & f_{10} = f^{aa_2a_3}\
   f^{aba_5}\ f^{ba_1a_4},\,f_{11}= f^{aa_1a_2}\
   f^{aba_5}\ f^{ba_3a_4},\,f_{12}= f^{aa_1a_5}\
   f^{aba_2}\ f^{ba_3a_4} 
\end{align}
and 
\begin{align}
    &e_1= 6 f_1-6 f_2+6 f_3-6 f_4+6 f_5+6 f_6-13 f_8-f_9+f_{10}-f_{11},\,\nonumber \\
    & e_2 = 4 \left(6 f_2+13
   f_8+f_9\right),\, \nonumber \\
   & e_3 = 4 \left(6 f_2-6 f_5-f_8+f_9+f_{11}-f_{12}\right)\,.
\end{align}
In the above equations, $d_{4f}^{a_1 a_2 a_3 a_4 a_5}$ is defined as \begin{align}
    d_{4f}^{a_1a_2 a_3 a_4 a_5} = \frac{1}{4!\,C_A}  \left[ \text{Tr}\left( T_A^{a} T_A^{a_1} T_A^{a_2} T_A^{a_3}\right) + \text{symmetric permutations}\right] f^{a a_4 a_5}\,,
\end{align}
where $(T_A^{a})_{bc} = -if^{a b c}$ and $C_A$ comes from eq.~\eqref{eq:leadingZgA}. The color structures $d_{4f}^{a b c d e}$ only contribute to five-loop splitting functions.
The above Feynman rules with five legs were also derived very recently~\cite{Falcioni:2024xav} within a different framework. The results here are presented in a different form, and we leave a detailed comparison to the future. The above results are also available in ancillary files.

\section{Conclusions}

In the renormalization of off-shell operator matrix elements with a twist-two operator insertion, the physical operators $O_q,\,O_g$ mix with generally unknown gauge-variant (GV) operators. In \cite{Gehrmann:2023ksf}, we introduced a framework to systematically derive the counterterm Feynman rules associated with these GV  operators for arbitrary spin $n$. In this work, we further apply this method to determine the Feynman rules for the leading GV operators with up to five legs. We obtained and presented these results in a compact form in equations~\eqref{eq:OBFeynmanrules}, \eqref{eq:OCFeynmanrules} and \eqref{eq:OAFeynmanrules}, which represent the main findings of this talk. To compute the counterterms for four-loop splitting functions, we will need to insert these Feynman rules into two-point off-shell matrix elements for up to three loops, which will be addressed in a future publication.

\begin{acknowledgments}
We would like to thank Vasily Sotnikov for useful discussions. This work is supported in part by the European Research Council (ERC) under the European Union’s research and innovation programme grant agreement 101019620 (ERC Advanced Grant TOPUP).
\end{acknowledgments}

\appendix

\bibliographystyle {JHEP}
\bibliography{LGV}

\end{document}